\documentclass[aps,pra,a4paper,10pt,twocolumn,nofootinbib]{revtex4-2} % eqsecnum
\usepackage[T1]{fontenc}
\usepackage{mathptmx}
\usepackage{datetime}
\usepackage{amsmath}
\usepackage{amsfonts}
\usepackage{amsfonts}
\usepackage{mathrsfs}
\usepackage[mathscr]{euscript}
\usepackage[dvips]{graphicx}
\usepackage{fancyhdr}
\usepackage{colordvi}
\usepackage{hyperref} %[hypertex=true]
\usepackage{epsfig}
\usepackage{color}
\usepackage{bm}
\usepackage{enumitem}

\def\overstrike#1#2{{\setbox0\hbox{$#2$}\hbox to \wd0{\hss
    $#1$\hss}\kern-\wd0\box0}}

\renewcommand{\Vec}{\bm}
\newdateformat{yymmdddate}{\THEYEAR/\twodigit{\THEMONTH}/\twodigit{\THEDAY}}

\def\UW{\mathscr{W}}

\begin{document}
\title{A multi-agent model of hierarchical decision dynamics}
\author{Paul Kinsler}
\homepage[]{https://orcid.org/0000-0001-5744-8146}
%\email{Dr.Paul.Kinsler@physics.org}
\affiliation{
  Department of Electronic \& Electrical Engineering, 
  University of Bath,
  Bath,
  BA2 7AY,
  United Kingdom
}

\lhead{\includegraphics[height=5mm,angle=0]{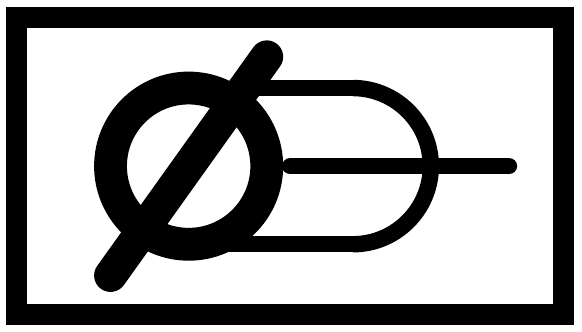}~~DECITREE}
\chead{Hierarchical decision dynamics}
\rhead{
\href{mailto:Dr.Paul.Kinsler@physics.org}{Dr.Paul.Kinsler@physics.org}\\
\href{http://www.kinsler.org/physics/}{http://www.kinsler.org/physics/}
}
%%\lfoot{\thesection . \thesubsection; ~~~~ (\yymmdddate\today:\currenttime) 
%% \href{http://localhost/}{(0)}}
%%\rfoot{{\large \emph{Not for redistribution}}}

\begin{abstract}

Decision making can be difficult when there are many actors
 (or agents)  
 who may be coordinating or competing 
 to achieve their various ideas of the optimum outcome.
Here I present a simple decision making model 
 with an explicitly hierarchical binary-tree structure, 
 and evaluate how this might cooperate to take actions
 that match its various evaluations of the 
 uncertain state of the world.
Key features of agent behaviour are (a) the separation of its
 decision making process into three distinct steps: 
 observation, 
 judgement,
 and action; 
 and (b) the evolution of coordination by the sharing of judgements.

\end{abstract}

%\pacs{...}

\date{\today}
\maketitle
\thispagestyle{fancy}

%
%\tableofcontents

\def\tinyzero{{\scalebox{0.6}{0}}}

\def\Aknow{\Phi}
\def\Akmin{\Aknow_{min}}
\def\Elink{L}

\def\Trate{\alpha}
\def\Rrisk{R}

%
%% =======================================================================
\section{Introduction}\label{S-context}

Decision making has always been a potentially complex problem, 
 and arguably never more so when
 there are many competing decision types to be made, 
 when they apply to different scopes and arenas, 
 when outcomes may be uncertain, 
 and when there are many actors --
 with different levels of authority --
 who may be coordinating or competing 
 to achieve their various ideas of the optimum outcome.

There is of course existing research in this area
 (e.g. decision making as briefly summarized in \cite{Gonzalez-FB-2017hf},
 and the advantages of hierarchies in \cite{Halevy-CG-2011opr}), 
 but the deliberately abstract model presented here was constructed 
 as a thought experiment 
 and entirely without reference
 to any existing literature.
Notably,
 the results from the model presented here
 do not focus on problem-solving or learning, 
 but on how inter-agent communication and preferences
 affect the behaviour or performance of the system as a whole, 
 how (or if) the system might converge to a final state
 in a static world, 
 and how well-matched that state is to agent preferences,
 agent or network metrics, 
 or to the world itself.
It could, 
 nevertheless, 
 be interesting to add in a feedback step enabling 
 the agents to adapt their preferences to better match the world.
However, 
 going as far as a formal representation --
 or remodelling -- 
 in terms of neural networks \cite{Lin-XLLW-2023nn} 
 is not my intent, 
 although the analogy is certainly an interesting one
 \footnote{I am even tempted to imagine a society
 as a kind of gas of interacting agents who take inputs
 from other agents and their surroundings and 
 then provide summarised ``opinion'' outputs to others 
 thus acting as a kind of weakly structured neural network 
 (or perhaps ``neural mesh''), but the structure is not
 set by a designed input-layer to output-layer architecture, 
 but merely biased by
 the association preferences of the constituent agents.}.

Here I attempt a simple model of a process, 
 focussing mainly on structurally hierachical organizations, 
 where different levels in the hierachy 
 are subject to different disadvantages and different time scales.
This is a simple agent based model, 
 specialised here to a binary tree structure, 
 and contains agents with a specific class of decision mechanisms.
Nevertheless, 
 within the decision mechanism used, 
 there is considerable freedom to adjust parameters, 
 and the decision function used could easily be made more 
 complicated or sophisticated.
Here, 
 however, 
 I restrict myself to nearly identical agents 
 using the same decision parameters.

One key feature is that the ``decision'' process is split into 
 three distinct steps: 
 information gathering, judgement formation, and action.
Notably, 
 any agent's judgement about a best action is not necessarily
 the same as the action taken, 
 since (e.g.) the preferred action might be altered --
 or even overridden --
 by the judgements of higher level agents.
The other key feature is that agents share only their judgements, 
 and not their observations about the world, 
 or their actions.

This paper contains:

\begin{description}

\item[-] 
 a tree model for the interactions 
  of a multi agent system which is explicitly hierarchical, 
  both in terms of connectivity and speed of action.

\item[-]
 Each agent takes into account the state of the 
 world it inhabits, 
 and that of its superiors and subordinates 
 and combines these ...

\item[-]
 to create (a) a judgement
  about what should ideally be done, 
  and (b) a potentially dissimilar action that it will take.

\item[-]
 This model is analysed on the basis of some
   ``reasonable'' assumptions for parameter values,  
   for a range of cases,
 and some results 
  showing how a network behaves on contact with 
  a world unmatched to its expectations
 are presented.

\item[-]
 In particular,
 role and relevance of differing success metrics are discussed.

\end{description}

\def\Ju{J}
\def\Ac{A}
\def\Wo{W}
\def\Qa{Q}

\def\St{\Vec{\omega}}

\def\JJ{\Vec{\sigma}}

\def\AA{\Vec{\alpha}}

\def\Lev{\lambda}

\def\Noise{\psi}
\def\Noisy{\eta}

\def\Hammer{\epsilon}

% ==================================================================
\section{Model}

\begin{figure}
\includegraphics[width=0.9\columnwidth]{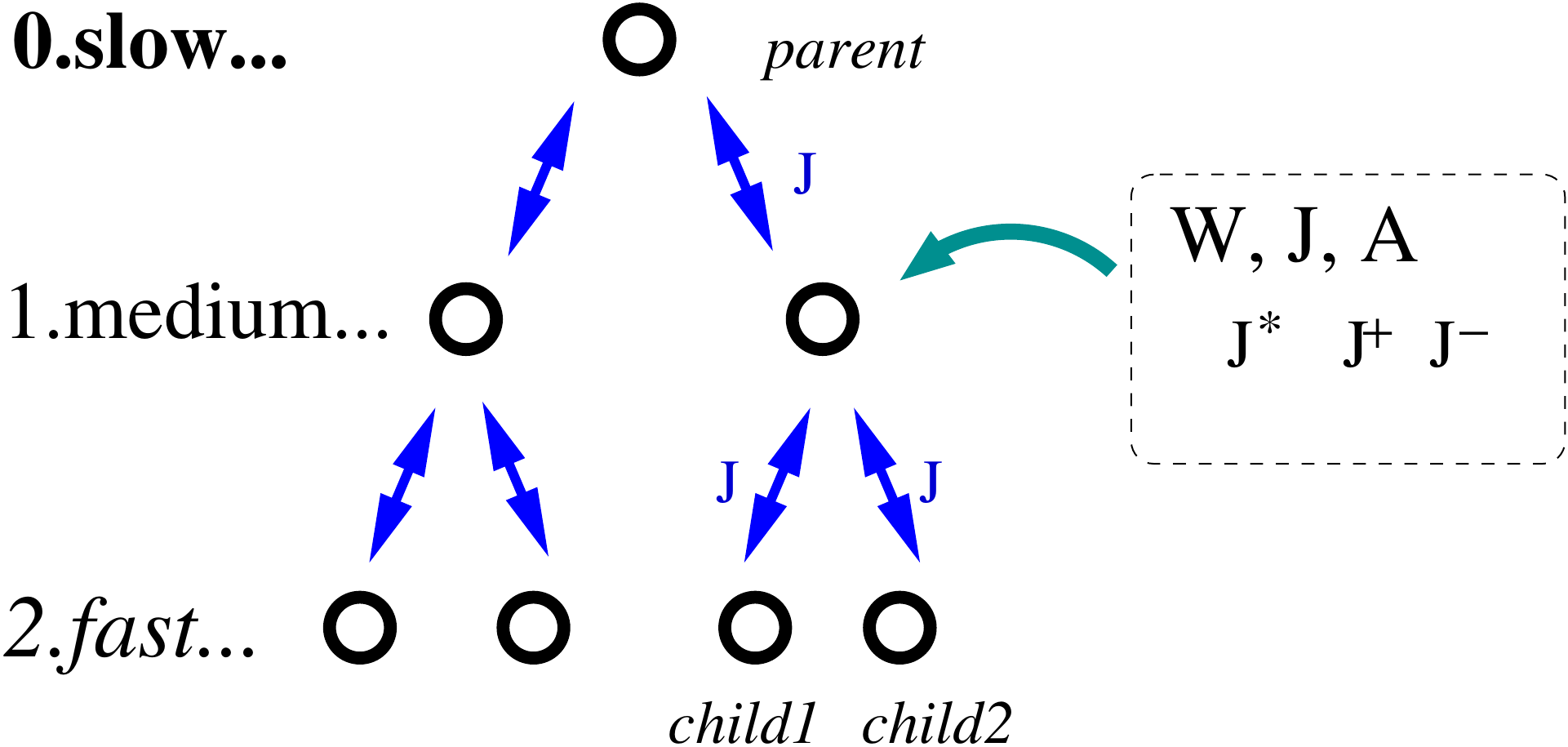}
\caption{A small three-level 
agent decision treee (agent-tree), 
 indicating the typical agent state values as described 
 in the main text, 
 the sharing of judgements between agents, 
 and the speed of operation 
 of the different levels.}
\label{fig-dtree}
\end{figure}

Here our decision hierarchy (network) is a binary tree
 containing $N=2^n$ agents.
This means the main interaction structure is of a parent agent
 with its two child agents; 
 although any agent might fulfill both parent and child roles.
As noted above, 
 for simplicity each agent $i$ behaves in the same way,
 and shares its \emph{judgement} ${\Ju}_i$
 about the best course of \emph{action} ${\Ac}_i$
 both up and down the tree.

The top level --
 the ``highest'' level -- 
 is level zero, 
 and the agent there is the ``ur-parent'', 
 and has no parent itself.
For convenience we give this ur-parent the index $i=0$.
The bottom level -- lowest level --
 ``leaf'' agents have parents, 
 but no children.
All other agents
 have both parent agents
 and child agents, 
 and an agent interacts only with its parent or its children.
A small example tree network is depicted
 on fig. \ref{fig-dtree}.

In what we present here we restrict all state, 
 judgement, 
 and action values
 to be real numbers.

Note that most of what is presented below generalises straightforwardly 
 to other types of tree-like agent networks, 
 whether binary or not, 
 or symmetric or not.
Even other non-tree networks might be modelled, 
 although some adaptions will need to be made --
 i.e. imposition of a ``level'' property on agents, 
 and reimagining the meaning of ``parent'' and ``child''.

% ------------------------------------------------------------------
\subsection{Notation}

We denote the judgement of an agent (with index) $i$
 at some time tick $t_j$ 
 to be ${\Ju}_i(t_j)$, 
 its action to be ${\Ac}_i(t_j)$; 
 these will be based in part on the agent's measurement ${\Wo}_i(t_j)$
 of the world state.
If we wish to refer to any of the possible elements 
 of agent state,
 we will use ${\Qa}_i(t_j)$; 
 i.e. $\Qa \in \{ \Ju, \Ac, \Wo \}$. 
However, 
 we will typically omit the time argument if it is not needed; 
 typically this is because all quantities considered will have the 
 \emph{same} time argument.

Since it will be necessary to refer to the parents and children 
 of an agent $i$, 
 we denote the index of that parent as $p(i)$, 
 and the indices of the children of $i$ as $u(i), v(i)$.
However,
 the top level agent (ur-parent) has no parent, 
 and the bottom level agents (leaves) have no children.
To address this we treat 
 the ur-parent as its own parent, 
 and the bottom level agents as their own children, 
 so that in those specific cases, 
 and only those cases
 we set $p(i)=i$, or set $u(i)=v(i)=i$.

Lastly, 
 since the level of the agent, 
 as determined by counting down the tree from
 the top (level 0), 
 we denote this as $\Lev(i)$.

% ------------------------------------------------------------------
\subsection{World}

We call the environment that our agents operate 
 the ``world'', 
 and in most of the work presented here this is 
 represented by
 a single real number $\UW$, 
 although an agent $i$'s measurement ${\Wo}_i$ of that state 
 may be inaccurate.

It is convenient to think of a state value of zero as 
 a situation requiring no action ($\UW=0$), 
 a judgement that no action is necessary (${\Ju}_i=0$), 
 or an agent taking no action (${\Ac}_i=0$); 
 and that values deviating from zero are somehow
 ``more extreme'' the larger the value.
If $\UW={\Ju}_i={\Ac}_i$
 then we should imagine that the judgement ${\Ju}_i$
 is exactly appropriate to the world situation $\UW$, 
 and that ${\Ju}_i$'s commensurate action ${\Ac}_i$ 
 is likewise also somehow perfectly matched to the situation.

It is simplest to assume that actions taken by the agents
 do not affect the world, 
 which means that any investigation focusses on 
 how well some specific agent-tree will adapt to 
 the world it inhabits.
However
 this is not a requirement, 
 so:\\
~

\noindent
\textbf{XW:} \emph{Optionally},
  the actions of each agent can "hammer" on the world and change it. 
 Each hammer blow (change) is proportional to that agent's (${\Ac}_i-\UW$), 
  divided by the number of agents,
 and multiplied by some prefactor $\Hammer$.

% ------------------------------------------------------------------
\subsection{Agent state}

Each node (agent) in the tree has three items of local state.
Here we consider the case where 
 each element of state is simply a single real number, 
 and if the values match they ``agree'' in some decision sense, 
 and any mismatch indicates a proportional degree of disagreement.
Specifically, 
 for any agent $i$
 these state elements consist of three local ones:

\begin{description}

\item[S1.]
  its observation ${\Wo}_i$ of the world value $\UW$,

\item[S2.]
   its (most recent) judgement ${\Ju}_i$ of what to do, which ideally would match $\UW$ exactly,

\item[S3.]
   its (most recent) action ${\Ac}_i$ taken, which ideally would (also) match $\UW$ exactly;

\end{description}

 as well as three state elements from its neighbours:

\begin{description}

\item[S4.]
   the most recent judgement of its parent ${\Ju}_i^* = {\Ju}_{p(i)}$,

\item[S5, S6.]
 the most recent judgements of its children ${\Ju}_i^\pm = {\Ju}_{u(i)}, {\Ju}_{v(i)}$,

\end{description}

An agent's state is thus contained in a six-vector:
~
\begin{align}
  {\St}_i
&=
  \left(
    {\Wo}_i, 
    {\Ju}_i, 
    {\Ac}_i, 
    {\Ju}_i^*, 
    {\Ju}_i^+, 
    {\Ju}_i^-
  \right)
\end{align}

Here we use the auxilliary definitions ${\Ju}_i^*,{\Ju}_i^+,{\Ju}_i^-$
 because any agent $i$ learns or gathers these at a particular time, 
 so they remain fixed until again updated; 
 whilst the parent or child agent may update their actual ${\Ju}$ values 
 independently.

% ------------------------------------------------------------------
\subsection{Agent algorithm}

Agents follow a three-step process,
 where 
 each step enables the agent $i$ to update its state vector ${\St}_i$
 in the sequence 

\begin{description}

\item[T1.]
    measure the world ($\UW$), 
     to obtain an updated value for ${\Wo}_i$, 
  and collect a new values for
  its parental judgement ${\Ju}_i^* = {\Ju}_{p(i)}$
  and child judgements ${\Ju}_i^+ = {\Ju}_{u(i)}$, ${\Ju}_i^- = {\Ju}_{v(i)}$, 
 and update ${\St}_i$;

\item[T2.]
    use its current state ${\St}_i$ to decide
 on a new judgement ${\Ju}_i'$, 
 and again update ${\St}_i$; 

\item[T3.]
    use its current state ${\St}_i$ to take
 its next action ${\Ac}_i'$,
 once again updating ${\St}_i$.

\end{description}

% ------------------------------------------------------------------
\subsection{Hierarchy}\label{SS-Hierarchy}

The binary tree model used here is already 
 hierarchical.
However, 
 in addition to this intrinsic behaviour
 we add two extra properties.

\begin{figure}
\includegraphics[width=0.9\columnwidth]{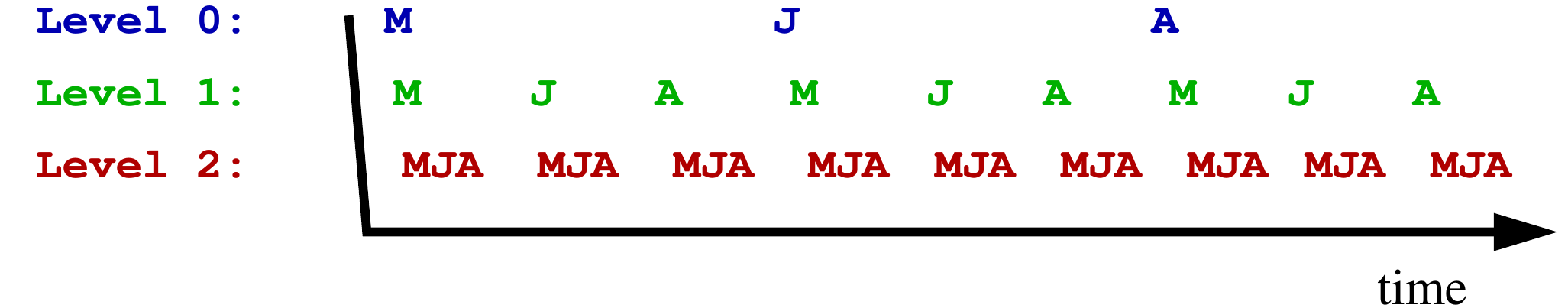}
\caption{Timing diagram:
 higher level agents act first, 
 but lower level agents act more frequently.
The step labelled M is the measurement step \textbf{T1}, 
 that labelled J is the judgement step  \textbf{T2},
 and 
 that labelled A is the action step  \textbf{T3}.
}
\label{fig-timing}
\end{figure}

First, 
 to incorporate the different timescales 
 desired for (slow) high level agents 
 and (faster) lower level agents,  
 for every one step a parent agent takes,
 its children complete all three steps (T1,T2,T3);
 and in any single time tick
 parents act before children \footnote{Arguably,
  this should be different, with e.g. the three steps 
   being treated as a unitary whole, and rather than the 1:3 
    ratio defined by the three steps, be some thing else, 
     perhaps 1:2.}
The relative timings of these steps 
 for agent of different levels are indicated
 in fig. \ref{fig-timing}.
Although alternative timing schemes might be imagined, 
 here we choose this one because of the interleaving
 of different types of steps taken at different levels.

\def\Noise{\psi}

Second, 
 we allow measurements of the world value $\UW$
 to have an accuracy that varies with 
 agent level, 
 as controlled by a parameter $\Noise$,
 a feature which will be addressed in detail below.

% ------------------------------------------------------------------
\subsection{Measurements, Judgements, and Actions}\label{SS-MeJuAct}

% - - - - - - - - - - - - - - - - - - - - - - - - - - - - - - - - - 
\subsubsection{T1: Measurements}\label{SS-Measurements}

A \emph{measurement} ${\Wo}_i$ of the world by an agent $i$
 returns the current (global) value $\UW$,
 plus some random noise; 
 i.e. 
~
\begin{align}
  {\Wo}_i &= \UW + \Noisy {\Noise}^{\Lev(i)} \xi
,
\end{align}
 where $\xi$ is a random variable with zero average $<\xi>=0$.

Here we imagine that in most cases, 
 lower level agents see proportionately more noise,
 reflecting (perhaps) the larger variation in their local conditions; 
 as opposed to the coarse-grained averaging
 expected of higher level decision makers.
Thus the amplitude of the added noise varies 
 from one level to the next lower according to a 
 proportionality constant $\Noise$; 
 e.g. at three levels down
 the relative noise multiplier is $\Noise^3$.
This scaling is chosen to make parent-child comparisons
 of noise contributions
 self similar
 between different levels.
Note that \emph{if} we preferred low level agents to (e.g.) 
  ``see more clearly'', 
 we could set $\Noise<1$, 
 adapt $\Noisy$ appropriately, 
 and have low level agents experience small measurement noise
 and high level agents seeing larger noise.

For any agent $i$, 
 collection of parental judgements 
  ${\Ju}^*={\Ju}_{p(i)}$
 and child judgements 
  ${\Ju}_i^+ = {\Ju}_{u(i)}$, ${\Ju}_i^- = {\Ju}_{v(i)}$
 is taken to be perfect and without error.

% - - - - - - - - - - - - - - - - - - - - - - - - - - - - - - - - - 
\subsubsection{T2: Judgements}\label{SS-Judgements}

A \emph{judgement} ${\Ju}_i$ is a weighted sum
 over an agent $i$'s six state elements. 
There are thus six possible judgement parameters. 
It is \emph{arguably reasonable} that any agent's judgement
 is most strongly influenced by its own ${\Wo}_i$, 
 but the model below requires no such restriction.

Here we specialise to a case where
 all the agents to have the same judgement weights,
 which will be defined to match the
 ``arguably reasonable'' idea above, 
 so that 
~
\begin{align}
  \JJ &= \left( 1-3\theta,0,0,\theta,\theta,\theta \right)
,
\label{eqn-JJ}
\end{align}
 where our default value will be
 $\theta=1/10$.
Note that this choice of $\JJ$ ignores
 an agents past judgements and actions,
 and is thus \emph{memoryless}; 
 and since it sums to unity can be coonsidered \emph{conservative}.

Here we will restrict ourselves to a simple linear
 computation of a judgement
~
\begin{align}
  {\Ju}_i' = {\St}_i \cdot {\JJ}_i
,
\end{align}
 which means that we then must update ${\St}_i$
 to incorporate the new ${\Ju}_i$ value by replacing the old.

% - - - - - - - - - - - - - - - - - - - - - - - - - - - - - - - - - 
\subsubsection{T3: Actions}\label{SS-Actions}

An \emph{action} $A$ is also a weighted sum of an agent $i$'s state. 
There are thus six possible action parameters. 
It is \emph{arguably reasonable} that any agent's actions
 are most strongly influenced by its parent's judgement, 
 but the model below requires no such restriction.

Here we specialise to a case where
 all the agents to have the same action weights,
 which will be defined to match the
 ``arguably reasonable'' actions above, 
 so that 
~
\begin{align}
   {\AA} &= \left( 0,\phi,0,1-\phi,0,0 \right)
,
\label{eqn-AA}
\end{align}
 where our default value will be
 $\phi = 2/10$.

Note that this choice of $\AA$ ignores
 an agents past judgements and actions,
 and is thus \emph{memoryless}; 
 and since it sums to unity can be considered \emph{conservative}.

Here we will restrict ourselves to a simple linear
 computation of the chosen action
~
\begin{align}
  {\Ac}_i' = {\St}_i \cdot \AA
\end{align}
 which means that we then must update ${\St}_i$
 to incorporate the new ${\Ac}_i$ value by replacing the old.

% ------------------------------------------------------------------
\subsection{Missing features}\label{SS-Missing}

\def\Fe{F}

A significant omission in this deliberately simple model 
 is any sense of logistics or action constraint; 
 which is problematic if 
 the world state is modified (``hammered'') by agent action.
Since we assume actions $\Ac$ are somehow more extreme 
 if they have larger values, 
 placing some limit on the rate of $\Ac$ expended would 
 seem reasonable; 
 and also mollify any tendency of a runaway to ever larger $\UW$
 by strategies that promote --
 by accident or design -- 
 increasingly aggressive agents.

Of course, 
 several suitable modifications suggest theselves, 
 but all require extra complexity in a model that aims at simplicity, 
 but nevertheless already has many parameters.
Therefore we defer treatment of this to later work.

Nevertheless,  
 perhaps the simplest logistics constraint would be to add 
 one extra element of agent state, 
 namely fuel ${\Fe}_i$.
This could be incremented at every time tick, 
 perhaps only up to some maximum ${\Fe}_{MAX}$.
Then
 an agent $i$ could only take some action ${\Ac}_i$ if ${\Fe}_i > {\Ac}_i$, 
 after which the fuel ${\Fe}_i$ would be decreased by ${\Ac}_i$.
Note that such extra detail should
 be an \emph{internal} feature of 
 some more sophisticated action determination
 (\ref{SS-Actions}), 
 rather than an explict additional part of the algorithm 
 proposed here.

% ==================================================================
\section{Initial conditions, Parameters, Metrics}\label{SS-InitParaMetric}

% - - - - - - - - - - - - - - - - - - - - - - - - - - - - - - - - - 
\subsubsection{Initial conditions}

Many possible initial conditions could be considered, 
 but here we start all agents 
 with $\Wo, \Ju, \Ac$ all zero, i.e. inactive. 
The world starts at some non-zero finite value (eg $\UW=3$), 
 so that the tree of agents has some discrepancy to adapt to.

% - - - - - - - - - - - - - - - - - - - - - - - - - - - - - - - - - 
\subsubsection{Parameters}

Here we choose our default measurement noise 
 proportionality to be $\Noise = \sqrt{2}$, 
 and the prefactor to be $\Noisy=10^{-3}$.
Here our random variable $\xi$ 
 is drawn from a rectangular distribution 
 spanning the range $[-1,+1]$.

If present (i.e. not zero),
 we choose the ``hammer'' parameter 
 to be ``small'' (e.g. $2 \times 10^{-3}$), 
 and scaled as $1/N$.
 (remember $N=2^n$).

The standard judgement parameters $\JJ(\theta)$ and 
 action parameters $\AA(\phi)$ 
 were given before in \eqref{eqn-JJ} and \eqref{eqn-AA}
 as part of the discusssion when they were introduced.

% - - - - - - - - - - - - - - - - - - - - - - - - - - - - - - - - - 
\subsubsection{Metrics}

Measures of success are here related to 
 the summed squares
 of differences between agent actions $\Ac$
 and some chosen state $\Qa$.
Thus for agents indexed by $i$, 
 we have success measure $X$ based on $\Qa$ as 
~
\begin{align}
  X_{\Qa} &= \sum_{i=0}^n \left( {\Ac}_i - {\Qa}_i \right)^2
.
\end{align}
Keeping in mind that the most success 
 matches the case when the measure sums to zero, 
 some possible measures that 
 might be considered are:

\begin{description}

\item[D0] naive success is when each agent's
 actions ${\Ac}_i$ match its observations, 
 i.e. when ${\Qa}_i={\Wo}_i$,

\item[D1] absolute success is when
 actions ${\Ac}_i$ match the world value, 
 i.e. when ${\Qa}_i=\UW$ --
 although of course no agent or set of agents
 could reliably calculate this,

\item[D2] perceived success is when 
 actions match judgements, 
 i.e. when ${\Qa}_i={\Ju}_i$,

\item[D3] bootlicker's success is when 
 actions match an agent's parent's judgement, 
 i.e. when ${\Qa}_i={\Ju}_i^*$,

\item[D4] authoritarian success is when 
 actions match the ur-parent's judgement ${\Ju}_0$

\item[D5] democratic success is when 
 actions match the agent-tree's average judgement 
 ${\Qa} = \bar{\Ju} = \sum_i {\Ju}_i$.

\end{description}

We can immediately see that D2 and D3 might be trivially 
 satisfied by appropriate choice of $\JJ$ or $\AA$.
However, 
 such choices will be unlikely to assist 
 if absolute success is also under consideration.

Other metrics also suggest themselves.
An agent tree in a malleable world may have a target value 
 of $\UW$ it wishes to achieve, 
 or many other possibilities.

An important point raised by such a selection of
 success criteria is how to balance the competing demands
 of several at once; 
 bearing in mind that the control parameters are 
 $\JJ$ and $\AA$, 
 and the results are combined (or mangled)
 by the hierarchical judgement sharing
 and the different timescales.

% ==================================================================
\section{Outcomes}\label{S-outcomes}

Here we assume \emph{arguably reasonable} agents, 
 as discussed in the preceeding section:
 those who base their judgements mostly on their observations, 
 but whose actions tend to follow orders from their superior, 
 as motivated by the preceeding discussion.
This behaviour is specified by our default parameters.

In the following figures, 
 the plotted quantities and axis scales
 are designed to normalise out the 
 gross effects of the tree size 
 (i.e. the number of agents $N$ in the tree).

% ------------------------------------------------------------------
\subsection{A clear and invariant World}\label{SS-noiselessnohammer}

The first case we look at will be for perfect vision
 (no observation noise) 
 and an unchanging world state $\UW$ (``no hammer'').

\begin{figure}
\includegraphics[width=0.89\columnwidth]{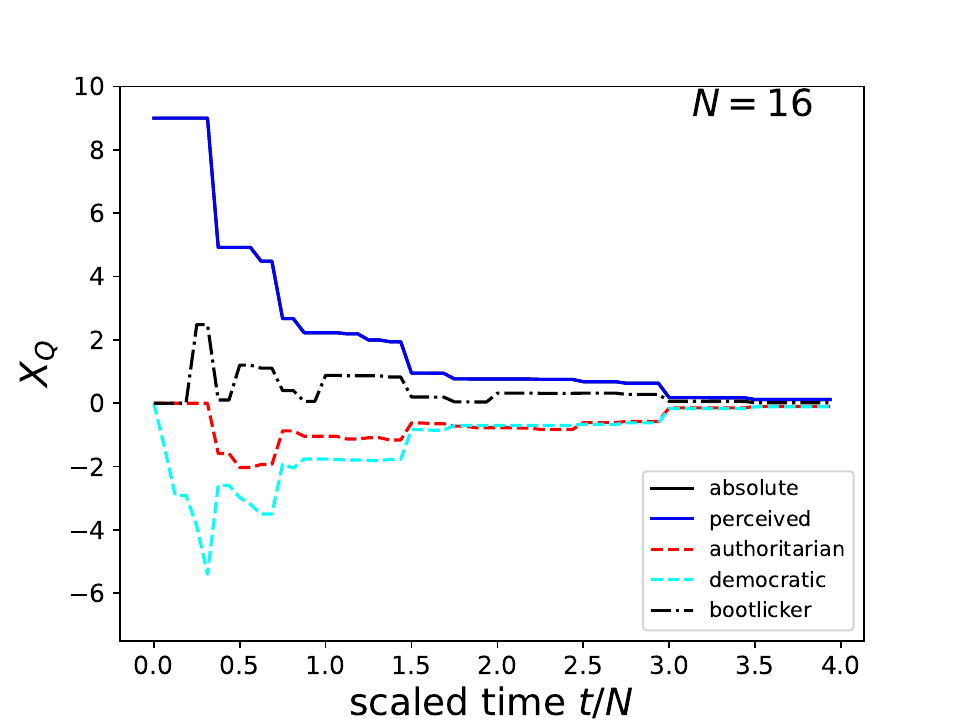}
\includegraphics[width=0.89\columnwidth]{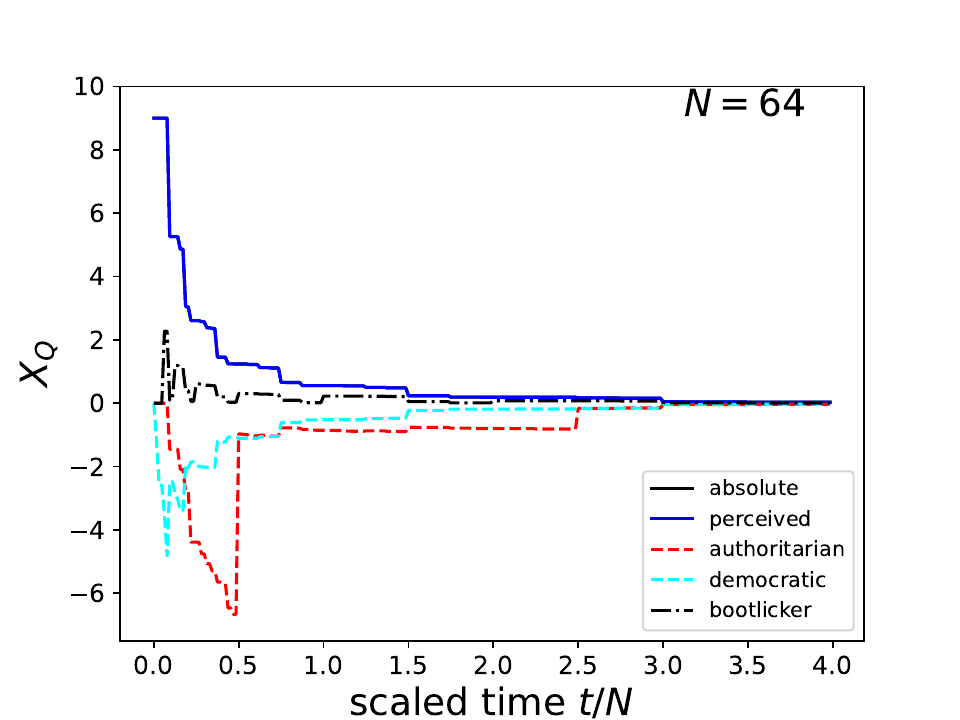}
\caption{Noiseless nohammer: 
 results for a small four-level tree with $N=16$
 compared to a larger six-level tree with $N=64$.
}
\label{fig-noiselessnohammer}
\end{figure}

As a consequence the behaviour of the system 
 is relatively striaghtforward -- 
 on fig. \ref{fig-noiselessnohammer}
 we see an essentially direct convergence to 
 high success for all metrics, 
 with $X_\UW$ and $X_\Ju$ being identical.
The structure (jumps and transitions) 
 seen on the results are deterministic, 
 and a result of distinct transitions
 caused by the algorithm and its interleaved timing.

% ------------------------------------------------------------------
\subsection{A foggy but invariant World}\label{SS-noisynohammer}

This second case introduces observation noise, 
 but still enforces an unchanging world state $\UW$.

\begin{figure}
\includegraphics[width=0.89\columnwidth]{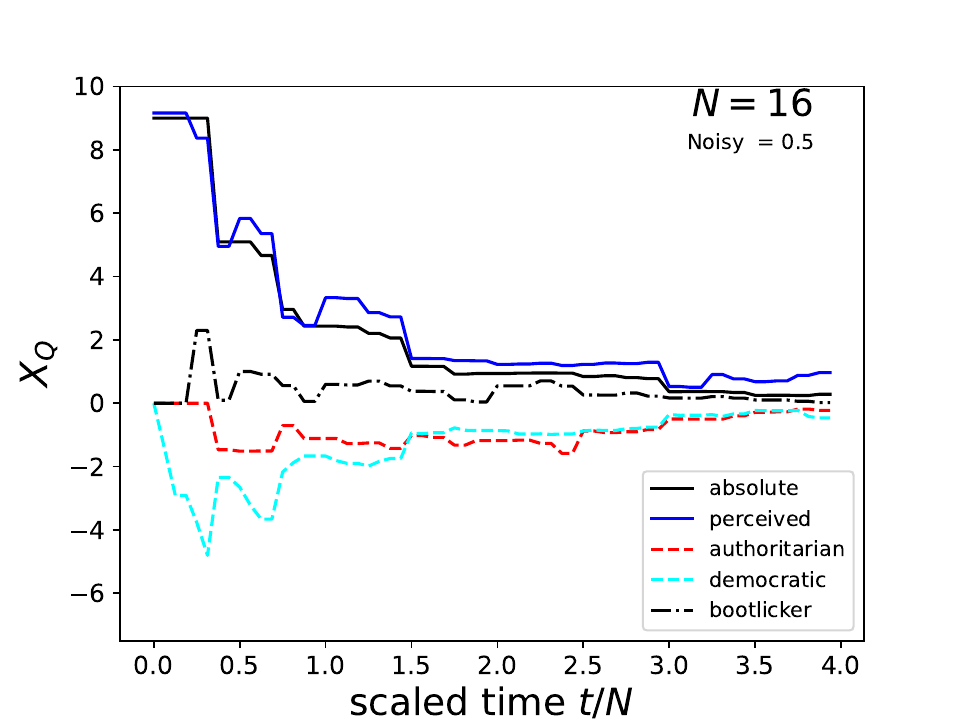}
\includegraphics[width=0.89\columnwidth]{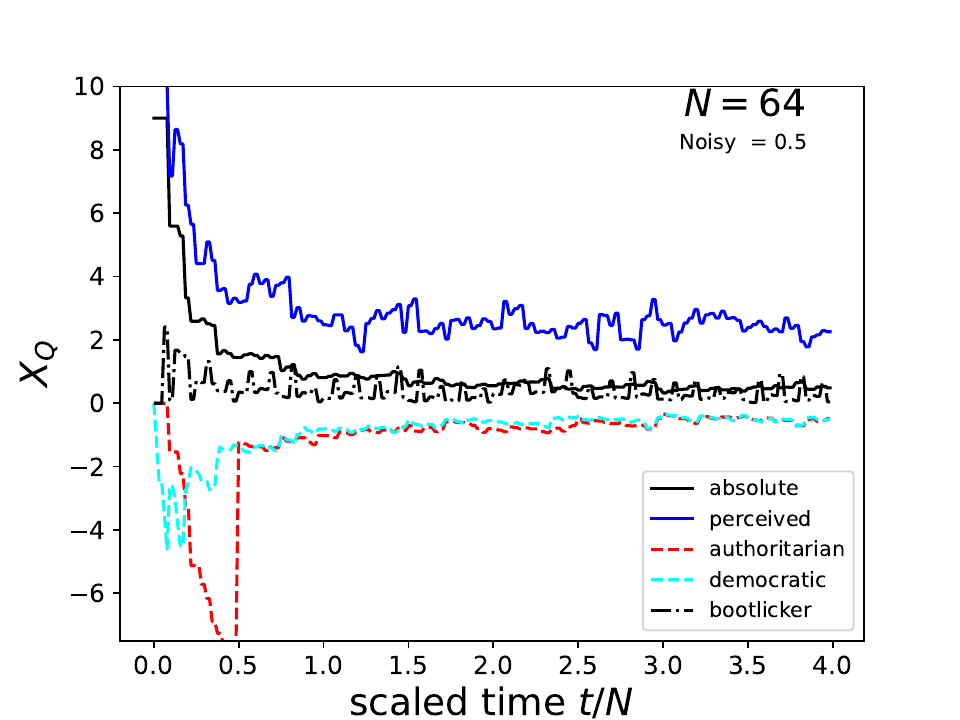}
\caption{Noisy nohammer: 
 results for a small four-level tree with $N=16$
 compared to a larger six-level tree with $N=64$.
The size of the observation noise parameter is
 indicated on the panels.
}
\label{fig-noisynohammer}
\end{figure}

In this case, 
 as shown on fig. \ref{fig-noisynohammer}, 
 we again we see a convergence
 to a state with excellent success metrics
 (i.e. zero valued).
However 
 we see an evolution
 that whilst similar to that on fig. \ref{fig-noiselessnohammer}, 
 is affected by noise, 
 and exhibits a moderate discrepancy between the network's
 absolute success $X_\UW$
 and its perceived success $X_\Ju$.

Since
 the perceived success does not converge to the absolute success; 
 it remains worse (and by by a non-negligible margin)
 than all the other measures shown.
This means that the agents will -- 
 on average --
 remain more dissatisfied with their performance than they should be.

% command line WNoisy ws 0.001
% command line Whammer_scale was 0e-3

% ------------------------------------------------------------------
\subsection{A clear but malleable World}\label{SS-noiselesshammer}

% command line Whammer_scale was 2e-3

\begin{figure}
\includegraphics[width=0.89\columnwidth]{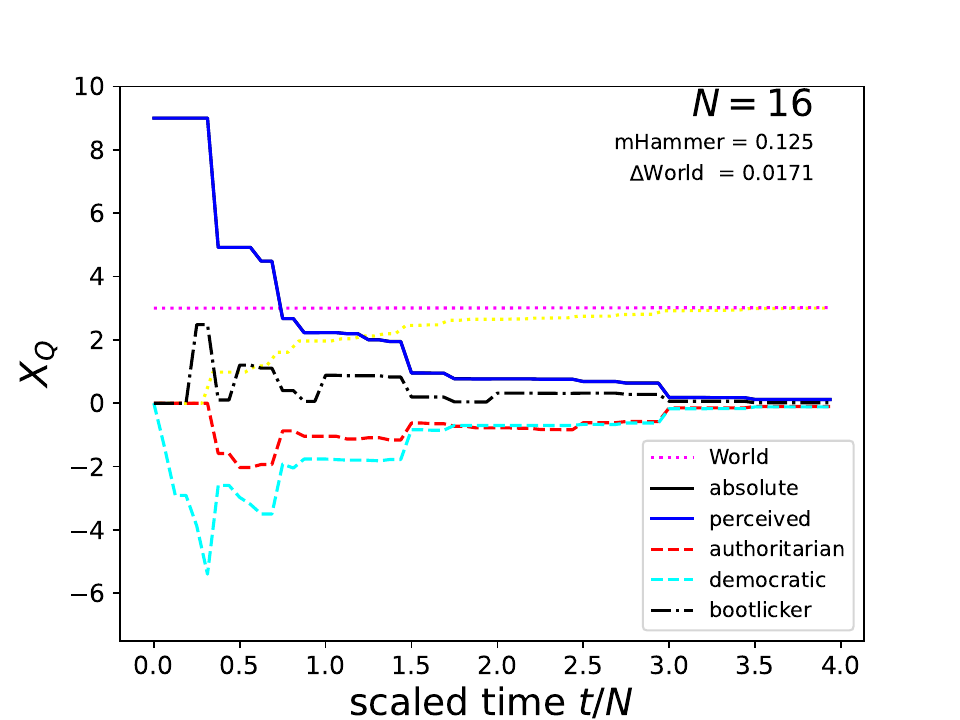}
\includegraphics[width=0.89\columnwidth]{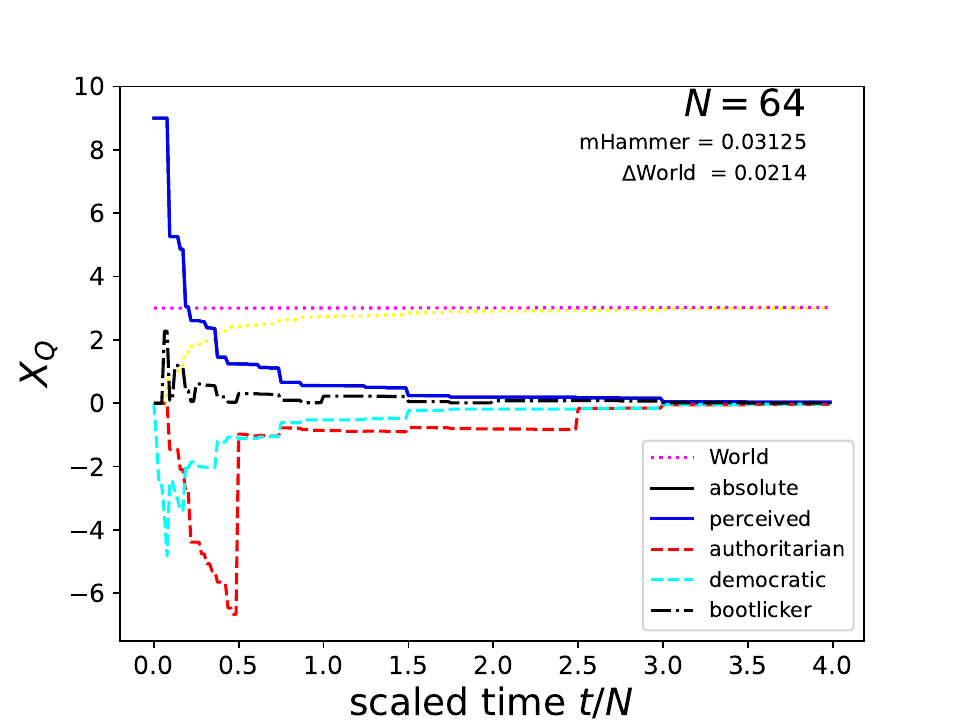}
\caption{Noiseless hammer: 
 results for a small four-level tree with $N=16$
 compared to a larger six-level tree with $N=64$.
The size of hammer parameter is
 indicated on the panels.
The near-horizontal dotted indicates the gradually increasing 
 value of the world state $\UW$; 
 the numerical size of the change is indicated 
 with the label $\Delta$World.
}
\label{fig-noiselesshammer}
\end{figure}

In the third case, 
 we incorporate only a 
 feedback (hammering) of agent actions
 on the world state.
Because I have chosen parameters 
 where the affect of the world is small, 
 here we see that fig. \ref{fig-noisyhammer}
 appears almost identical to fig. \ref{fig-noiselessnohammer};
 but there is a key difference.

Close examination, 
 and as recorded by the $\Delta$World value on the panels, 
 the world state value $\UW$ is slowly 
 but inexorably increasing with time.
This is indicated by the slight gradient 
 on the near-horizontal dotted line.
This runaway escalation in decision making behaviour 
 can be reduced or enhanced by altering the simulation parameters; 
 here they are tuned to keep the increase gradual.

Note that trying to moderate the escalation by 
 using non-conservative judgment or action weights
 (i.e. where $|\JJ|$ or $|\AA|$ are scaled to less than unity)
 is surprisingly ineffective.
This is likely because any mismatch between $\Wo$, $\Ju$, and $\Ac$
 will always tend to make things worse, 
 but it deserves closer investigation.

% ------------------------------------------------------------------
\subsection{A foggy and malleable World}\label{SS-noisyhammer}

% command line Whammer_scale was 2e-3

\begin{figure}
\includegraphics[width=0.89\columnwidth]{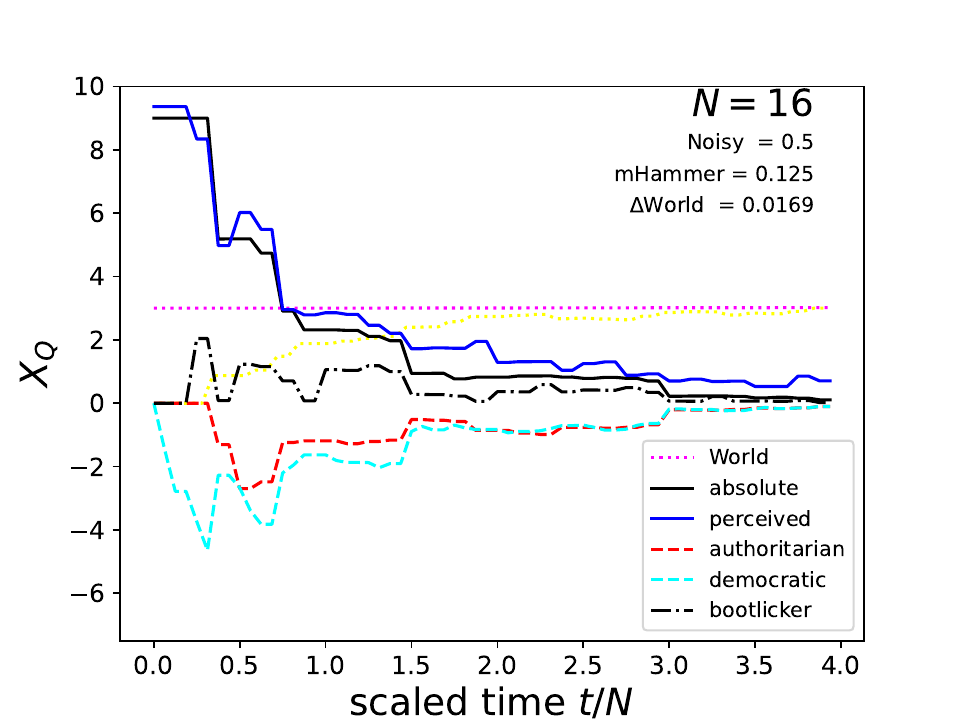}
\includegraphics[width=0.89\columnwidth]{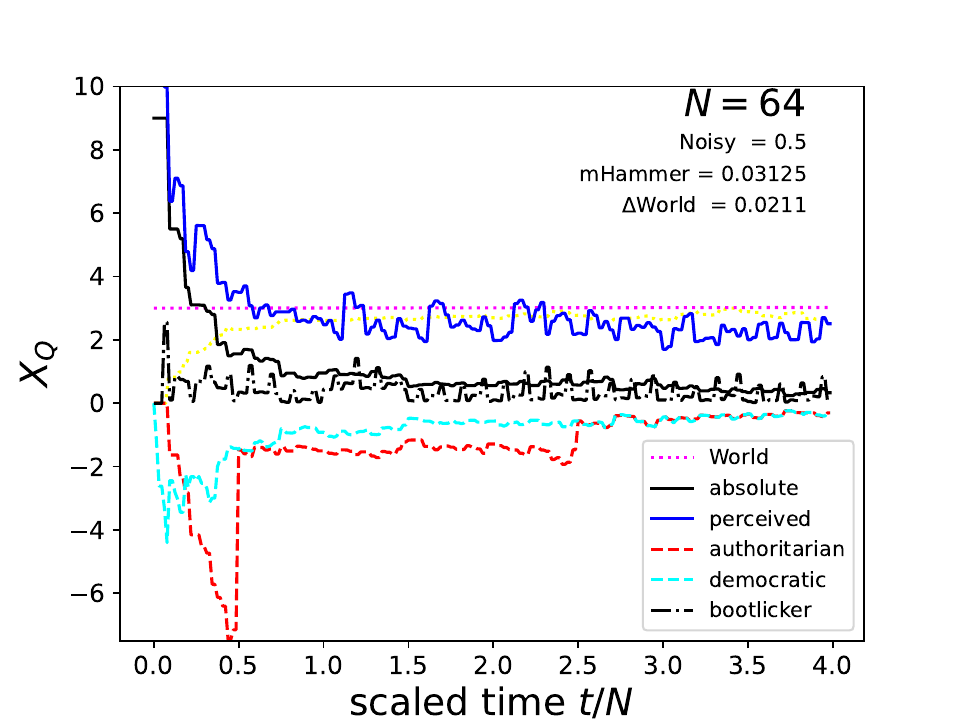}
\caption{Noisy hammer: 
 results for a small four-level tree with $N=16$
 compared to a larger six-level tree with $N=64$.
The size of the observation noise
 and hammer parameters are
 indicated on the panels.
The near-horizontal dotted indicates the gradually increasing 
 value of the world state $\UW$; 
 the numerical size of the change is indicated 
 with the label $\Delta$World.
}
\label{fig-noisyhammer}
\end{figure}

In the fourth and final case
 we incorporate both observation noise and
 feedback (hammering) of agent actions
 on the world state.

On first sight of fig. \ref{fig-noisyhammer},
 these are perhaps not dissimilar to a noisier version of the 
 ``Noisy Nohammer'' results of \ref{SS-noisynohammer}, 
 but there is the same key difference as for the noiseless hammer
 results presented in Sec. \ref{SS-noiselesshammer}.

First, 
 however, 
 we see again that
 the network's perceived success does not converge to the absolute success; 
 it remains worse than all the other measures shown, 
 although the margin is unchanged by the new ``hammer'' feature in the model.
Again, 
 the agents
 remain more dissatisfied with their performance than they should be.

Second,
 again we see world state value $\UW$ slowly 
 but inexorably increasing with time, 
 as indicated by the positive valued $\Delta$World number.

\newpage

% ==================================================================
\section{Future possibilities}\label{S-future}

There is obviously a great deal of scope --
 even within the model as presented here --
 to investigate a wider range of parameters.
Less conservatively, 
 there are also many extensions, 
 such as adaptive agents, 
 allowing diversity of agent properties,
 or more complicated and even dynamic judgement algorithms.

Nevertheless, 
 between these two extremes some interesting possibilities
 suggest themselves; 
 and would be interesting to pursue 
 once a classification of the 
 judgement and action parameter $\St, \AA$
 state space has been made.
Some possibilities are:

\begin{description}

\item[Changing world]
 Set the world $\UW$ to have an oscillation, 
 and investigate the interplay of the decision time scales 
 against the world's timescales.
The more unsympathetic of us might instead 
 test their network against a more complicated (or even chaotic)
 variation in $\UW$.

\item[Competing networks] 
 Have a pair of networks side by side, 
  and have the agents hammer on their counterpart --
  so we have that network I's agent action $A$ values
  being network II's agents world $W$ values,
 and vice versa. 
You might also try putting the networks foot to foot, 
 so that only the lowest level agents see each other.

\item[Strategic vs tactical]
 Have a world function where high level agents 
 see different $\Wo$ values to low level ones
 (perhaps even with opposite sign).
 In such a circumstance, 
 how might a network come to a sense of agreed action?

\end{description}

% ==================================================================
\section{Summary}\label{S-summary}

This hierarchical decision model is deliberately simple, 
 but attempts to incorporate 
 the essential features of 
 hierarchy (multiple levels),
 different decision time-scales, 
 the variation in perceptions, 
 and inter-level communications.

Thus, 
 in its favour, 
 it incorporates those design features, 
 whilst also having 
 configurable (and generalisable) judgement and action steps.
Further, 
 since it is the behaviour of an agent within the network which 
 is the core algorithmic feature, 
 it might be generalised to
 different sorts of trees or other network configurations.

However,
 the model does have some limitations:
 notably, 
 that the decisions themselves are handled 
 in a very abstract way, 
 and are all of similar type; 
 there is no mechanism to make high-level 
 (which we might want to consider ``strategic'') 
 functionally different to the mid- 
 or ``tactical'' low-level ones.
Further, 
 and in common with many abstract models, 
 whilst one might add more features 
 or improved algorithms, 
 such added complexity will not obviously make the model 
 more realistsic, 
 but may only obscure interesting generic behaviours.

The selection of success metrics proposed here, 
 and their explicit comparison,
 is  key to the understanding of evaluating decision processes:
 the same decision (or decision process) can have outcomes that 
 are regarded differently by different actors or agents --
 even those who are on the same side.
Indeed, 
 in several cases above we saw that the perceived success 
 can quite easily be less than the \emph{actual} success.

% ==================================================================
\acknowledgments

The author would like to acknowledge the role of the Maths of EMA
 workshop run by PA Consulting in early 2023.
 One of the topics discussed was relevant to this field,
 and spurred me to construct this model, 
 and to do this brief investigation.


%\bibliography{bibtex.bib}

\end{document}